\documentclass[preprint,english,aps,prl,epsfig,graphics]{revtex4}
\usepackage{ae}
\usepackage{aecompl}
\usepackage[T1]{fontenc}
\usepackage[latin1]{inputenc}
\usepackage{amsmath}
\usepackage{graphicx}
\usepackage{amssymb}

\makeatletter
\usepackage{graphicx}
\usepackage{amsmath}

\usepackage{babel}
\makeatother
\begin{document}

\title{Correcting low-frequency noise with continuous measurement}

\author{L. Tian}

\affiliation{National Institute of Standards and Technology, 100 Bureau Drive,
Stop 8423, Gaithersburg, MD 20878, USA }

\date{\today {}}

\begin{abstract}
Low-frequency noise presents a serious source of decoherence in solid-state
qubits. When combined with a continuous weak measurement of the eigenstates, the low-frequency noise induces a second-order relaxation
between the qubit states. Here we show that the relaxation provides
a unique approach to calibrate the low-frequency noise in the 
time-domain. By encoding one qubit with two physical qubits that are alternatively
calibrated, quantum logic gates with high fidelity can be performed.
\end{abstract}
\maketitle
Solid-state quantum devices have been demonstrated to be promising
systems for quantum information processing \cite{NielsenBook}. A key
factor that affects the scalability of such devices is decoherence \cite{Gottesman,KnillNature2005}.
The low-frequency noise \cite{WeissmanPMP1988} was shown to be a serious
source of decoherence and is ubiquitous in solid-state systems such
as single-electron tunneling devices and superconducting qubits \cite{SimmondsPRL2004,WellstoodAPL2004,vanHarlingenPRB2004,AstafievPRL2004,ZorinPRB1996}
and quantum optical systems \cite{Wineland2003}. Numerous theoretical
and experimental works were devoted to the study of the physical origin
of the low-frequency noise \cite{TheoryNoise}. It has been widely
accepted that the low-frequency noise features a spectrum of $1/f^{\alpha}$
with $\alpha\in[1,2]$ and a finite band width. Recently, several
approaches were studied to control the decoherence from low-frequency
noise, including the dynamical control technique \cite{NakamuraPRL2002,CollinPRL2004,ShiokawaPRA2004,GutmannPRA2005,FalciPRA2004,FaoroPRL2004}
and circuit designs at the degenaracy point \cite{VionScience2002,MakhlinPRL2004}. 

In the degeneracy point approach \cite{VionScience2002}, the low-frequency
noise only induces off-diagonal coupling which is much weaker than
the energy separation between the qubit states. Because of its low-frequency 
nature and the large energy separation, this coupling can
not generate transition between the qubit states. Decoherence of the
qubit is due to the second-order coupling between the noise and the qubit
and is significantly suppressed \cite{MakhlinPRL2004}. This scenario,
however, is changed when the qubit is continuously monitored by a
detector. The measurement, together with the low-frequency noise,
generates non-trivial dynamics for the qubit. 

Continuous measurement is a useful tool for studying stochastic quantum
evolution and was first explored in the context of quantum optics \cite{GardinerBook}.
Well-known applications of this technique include homodyne detection
of an optical field in a cavity and measurement of the small displacement
of a quantum harmonic oscillator \cite{CavesPRA1987,WisemanPRL1993,DohertyPRA1999,OreshkovPRL2005}.
Recently, continuous measurement was adapted to solid-state systems
such as quantum point contact and single-electron transistor to study
the effect of detector's back action on a qubit \cite{KorotkovPRB2001,MeasurementQubit}.
During a continuous measurement, the detector couples weakly with
the measured system and only slightly perturbs the system. The measurement
record contents a large contribution from the back action noise of
the detector and a small contribution from the measured system. Meanwhile,
the back action noise modifies the intrinsic dynamics of the measured
system and can result in interesting phenomena.

In this paper, we study the dynamics of a continuously monitored qubit
at the degeneracy point \cite{VionScience2002,MakhlinPRL2004}, where
the detector measures the eigenstates of the qubit. Assisted by the
measurement, the off-diagonal coupling of the low-frequency noise
induces relaxation between the qubit states in addition to dephasing.
It can be shown that the rate of the relaxation depends on the rate
of the measurement linearly and depends on the magnitude of the noise
quadratically. This stochastic process can be studied numerically
in the quantum Bayesian formalism \cite{KorotkovPRB2001} and
a real-time characterization of the low-frequency noise can
be achieved from the measured switchings between the eigenstates of
the qubit. We show that by using the measurement record to calibrate
the noise, high fidelity quantum logic operations can be performed.
This presents a novel approach to suppress the decoherence of the qubit
due to the low-frequency noise. 

Consider a qubit with an energy separation $2\hbar E_{z}$ between
the states $|0\rangle$ and $|1\rangle$. At the degeneracy point,
the noise induces off-diagonal coupling $\hbar\delta V$. The qubit
Hamiltonian is \begin{equation}
H_{0}=\hbar\left(\begin{array}{cc}
-E_{z} & \delta V\\
\delta V & E_{z}\end{array}\right).\label{eq:H0}\end{equation}
 For low-frequency noise with $\langle|\frac{\partial\delta V}{\partial t}/\delta V|\rangle\ll2E_{z}$
and $|\delta V|\ll2E_{z}$, the energy separation prevents the transition
between the qubit states and partially protects the qubit from decoherence.
The noise causes dephasing by a second-order coupling
as $H_{0}\approx\hbar(E_{z}+\delta V^{2}/4E_{z})\sigma_{z}$ with
a noise spectrum $\left(\frac{\delta V^{2}}{4E_{z}}\right)_{\omega}^{2}$ \cite{MakhlinPRL2004}.
Here $\sigma_{z,x}$ are the Pauli matrices of the qubit. 

A measurement of the eigenstates of the qubit is performed with the
detector current $I_{0}$ for the state $|0\rangle$ and $I_{1}$
for the state $|1\rangle$. The detector noise is described by a white
noise spectrum $S_{I}$ which determines the measurement rate $\Gamma_{m}=\Delta I^{2}/4S_{I}$,
the rate at which the qubit states can be resolved by the measurement,
with $\Delta I=|I_{0}-I_{1}|$. For a continuous measurement, we have 
$\Delta I\ll(I_{0}+I_{1})/2$ and $\Gamma_{m}\ll\sqrt{E_{z}^{2}+\delta V^{2}}$.
Hence, the measurement only slightly perturbs the qubit on the time
scale of $1/2E_{z}$.

By averaging over all the trajectories of the continuous measurement,
the master equation of the qubit can be written as \cite{GardinerBook}\begin{equation}
\frac{\partial\bar{\rho}}{\partial t}=-\frac{i}{\hbar}[H_{0},\,\bar{\rho}]-\frac{\Gamma_{m}}{4}[\sigma_{z},\,[\sigma_{z},\,\bar{\rho}]],\label{eq:rho-ave}\end{equation}
 where $\bar{\rho}$ is the density matrix with the ensemble average.
The last term in Eq.(\ref{eq:rho-ave}) describes the dephasing of the qubit 
due to the detector's back action noise. The low-frequency noise can
be treated as a perturbation in the above master equation. To second-order
approximation, the evolution of the qubit can be derived and the probability
of the state $|0\rangle$ follows $\partial\bar{\rho}_{00}/\partial t=-(\bar{\rho}_{00}-1/2)/\tau_{a}$
with a decay rate\begin{equation}
\tau_{a}^{-1}=\frac{4\delta V^{2}\Gamma_{m}}{4E_{z}^{2}+\Gamma_{m}^{2}}\label{rho_approx}\end{equation}
proportional to the measurement rate. The stationary state of the
qubit for $t\gg\tau_{a}$ has $\bar{\rho}_{00}^{s}=\bar{\rho}_{11}^{s}=1/2$ and $\bar{\rho}_{01}^{s}=\bar{\rho}_{10}^{s}=0$  which is a mixture of the states $|0\rangle$ and $|1\rangle$ with
equal probability. The information of
the initial state of the qubit is lost because of the measurement.
Hence, starting from an initial state $|0\rangle$, a switching to
the state $|1\rangle$ can occur in the presence of the continuous
measurement. Without the measurement at $\Gamma_{m}=0$, $\tau_{a}^{-1}=0$
and the switching can not occur. Note that in the limit of strong
coupling between the detector and the qubit with $\Gamma_{m}\gg E_{z}$,
the qubit dynamics is dominated by the quantum Zeno effect where fast
measurement prevents the relaxation of the qubit. 

In reality,  selective quantum trajectories determined by the stochastic
quantum master equation are recorded instead of the ensemble average.
For the selective process, the switching between the qubit states
occurs with the rate $\tau_{jp}^{-1}=\tau_{a}^{-1}/2$. Here the reduction
of the rate by a factor $1/2$ from $\tau_{a}$ 
comes from the equal probability of the states $|0\rangle$ and $|1\rangle$
after a sufficient long time of measurement. Hence,  monitoring
the switchings of the qubit continuously provides a time-domain characterization
of the low-frequency noise. 

We numerically simulate the selective process in the quantum
Bayesian formalism. Compared with the Ito and the Stratonovich formalisms,
the Bayesian formalism not only gives intuitive explanation of the
process, but also has better convergence property\cite{GardinerBook}. The time evolution of the continuous measurement is described by \cite{KorotkovPRB2001} \begin{align}
\frac{\rho_{00}(t+\delta t)}{\rho_{11}(t+\delta t)} & =\frac{\rho_{00}(t)e^{-(I(t)-I_{0})^{2}\delta t/S_{I}}}{\rho_{11}(t)e^{-(I(t)-I_{1})^{2}\delta t/S_{I}}}\nonumber \\
\rho_{01}(t+\delta t) & =\rho_{01}(t)\sqrt{\frac{\rho_{00}(t+\delta t)\rho_{11}(t+\delta t)}{\rho_{00}(t)\rho_{11}(t)}},  \label{eq:select-eq}\end{align}
where the current $I(t)$ is recorded at time $t$ over a short interval
$\delta t \ll 1/\Gamma_{m}$.  The measurement
record includes the average current of the qubit as well as the detector
noise:  $I(t)=I_{0}\rho_{00}+I_{1}\rho_{11}+\xi$, where $\xi$ is the
Gaussian noise of the detector with the spectrum $\langle\xi^{2}\rangle=S_{I}/2\delta t$.
The dynamics of the qubit during $\delta t$ is completed
by the Hamiltonian evolution: $\rho(t+\delta t)=\rho(t)-\frac{i}{\hbar}[H_{0},\rho]\delta t$.
Note that during the selective evolution, the qubit state is pure or purified. 
In the simulation, without loss of generality, we treat the low-frequency
noise as a Gaussian noise with $1/f$ spectrum. With discrete
frequency components separated by $\Delta\omega$ and bounded above
by the band width $B_{w}=N\Delta\omega$, we have \cite{GaussianNoise}
\begin{equation}
\delta V(t)=\sum_{i=1}^{N}\sqrt{\beta/\omega_{n}}\alpha_{n}\cos\left(\omega_{n}t+\varphi_{n}\right),\label{eq:dv}\end{equation}
 where $\alpha_{n}$ are Gaussian random numbers with $\langle\alpha_{n}^{2}\rangle=1$,
$\varphi_{n}$ are random numbers equally distributed between $0$
and $2\pi$, and $\beta$ is the magnitude characterizing the noisewith
the noise spectrum $\delta V_{\omega}^{2}=\beta/12\omega\Delta\omega$. 

\begin{figure}
\includegraphics[%
  bb=60bp 240bp 567bp 612bp,
  clip,
  width=7.5cm]{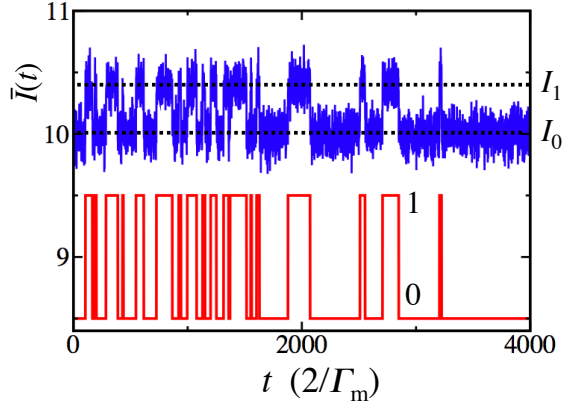}

\caption{\label{cap:fig1}Numerical simulation of the selective evolution
at $\delta V(0)=0.82$. Upper panel: the average current $\bar{I}(t)$;
lower panel: the filtered data correponding to the states $|0\rangle$
and $|1\rangle$.}
\end{figure}

Starting from the initial state $|0\rangle$, the selective evolution
is first simulated over a duration $2n_{p}/\Gamma_{m}$ with $n_{p}=2000$. For
$\delta t\ll1/\Gamma_{m}$, the measured current $I(t)$
contents a large fluctuation with $\sqrt{\langle\xi^{2}\rangle}\gg\Delta I$.
We process the measurement record by averaging the current over an
interval $2/\Gamma_{m}$: $\bar{I}(2n/\Gamma_{m})=\frac{\Gamma_{m}\delta t}{2}\sum I(t)$
for $2n/\Gamma_{m}\le t<2(n+1)/\Gamma_{m}$. The average current $\bar{I}$
now contents a fluctuation smaller than $\Delta I$ and can be used
to indicate the qubit states. In Fig.\ref{cap:fig1}, $\bar{I}$ is
plotted versus time with $\bar{I}(0)\approx I_{0}$. During the evolution,
the average current switches between $\bar{I}\approx I_{0}$ and $\bar{I}\approx I_{1}$
corresponding to switchings between the states $|0\rangle$ and $|1\rangle$.
The number of switchings is determined by the rate $\tau_{jp}^{-1}$, from which
we derive the first estimation of the noise
magnitude $\delta V_{1}\approx|\delta V|$. To determine the sign
of the noise, we shift the off-diagonal coupling in Eq. (\ref{eq:H0})
by a constant $-\hbar\delta V_{1}/2$ and continue the simulation
for another duration $2n_{p}/\Gamma_{m}$. For $\delta V>0$, the total 
off-diagonal coupling is decreased after the shift and subsequently the
number of switchings is decreased in the second part of the simulation.
For $\delta V<0$, the total off-diagonal coupling is increased after
the shift and so does the number of switchings. After the evolution,
we derive the second estimation of the noise magnitude $\delta V_{2}$.
It can be shown that $\delta V_{2}=|\delta V|/2$ for $\delta V>0$ and
$\delta V_{2}=3|\delta V|/2$ for $\delta V<0$. Finally, 
the noise is estimated as \[
\delta V_{c}=\{\begin{array}{cc}
\frac{\delta V_{1}}{2}+\delta V_{2}, & \textrm{for}\,\delta V_{1}>\delta V_{2}\\
-\frac{\delta V_{1}}{2}-\frac{\delta V_{2}}{3}, & \textrm{for}\,\delta V_{1}<\delta V_{2}\end{array}\]
with both magnitude and sign. The parameters used in Fig.\ref{cap:fig1}
are $E_{z}=7$, $\Gamma_{m}=0.1$, $\delta V(0)=0.82$ at the start of the simulation, $\delta V(4n_{p}/\Gamma_{m})=0.87$ at the end of the simulation, $I_{0}=10$ and $I_{1}=10.4$ with
the noise spectrum $S_{I}=0.4$. To count the number of switchings, the
current $\bar{I}$ is filtered to be $0$ or $1$, which is also shown
in Fig.\ref{cap:fig1}. In the first part of the simulation, the
number of switchings is $n_{jp}^{(1)}=29$; in the second part,
the number of switchings is $n_{jp}^{(2)}=7$. It is obvious from
Fig. \ref{cap:fig1} that the switchings in the second part are
much sparser than that in the first part. The estimated noise is
$\delta V_{c}=0.73$. By shifting the off-diagonal coupling
in Eq. (\ref{eq:H0}) with the calibrated value $\delta V_{c}$, the
dephasing can be reduced by a factor $\left|\frac{\delta V_{c}-\delta V(4n_{p}/\Gamma_{m})}{\delta V(0)}\right|^{4}\approx10^{-3}$.

\begin{figure}
\includegraphics[%
  bb=66bp 312bp 552bp 660bp,
  clip,
  width=7.2cm]{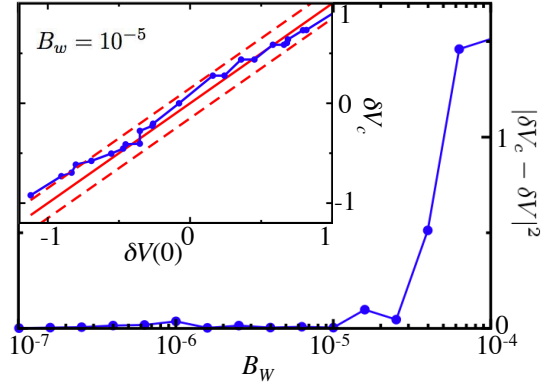}
\caption{\label{cap:fig2}Main: residue noise $|\delta V_{c}-\delta V(4n_{p}/\Gamma_{m})|^{2}$
versus band width $B_{w}$. The solid circles connected by the solid
curve are the residue noise from the simulation. Inset: $\delta V_{c}$
versus $\delta V(0)$ at $B_{w}=10^{-5}$. The solid circles connected
by the solid curve are $\delta V_{c}$. The straight solid line in
the middle is $\delta V(0)$ and the two dashes lines are $\delta V(0)\pm0.15$.}
\end{figure}

The effectiveness of the above calibration depends sensitively on
the band width of the low-frequency noise. In Fig.\ref{cap:fig2},
the average squared residue noise $\langle|\delta V_{c}-\delta V(4n_{p}/\Gamma_{m})|^{2}\rangle$
is plotted versus the band width $B_{w}$. For $B_{w}<(4n_{p}/\Gamma_{m})^{-1}$,
the residue noise is much smaller than $\langle\delta V^{2}\rangle$.
For $B_{w}\ge(4n_{p}/\Gamma_{m})^{-1}$, the residue noise increases
dramatically. Thus, when the duration of the calibration is shorter
than the time scale of the noise, the calibration is successful. With
our parameters, $(4n_{p}/\Gamma_{m})^{-1}=1.3\cdot10^{-5}$ agrees
well with the sharp increase in the residue noise near this band width.
In the inset of Fig.\ref{cap:fig2}, the calibrated value $\delta V_{c}$
is plotted versus the initial value of the noise at $B_{w}=10^{-5}$.
It can be seen that the residue noise is mostly
limited within $\pm0.15$. 

The residue noise of the calibration results from two factors: the
finite accuracy due to finite number of switchings during the continuous
measurement, and the time evolution of the low-frequency noise during
the measurement. First, the probability of switching during a time
$t$ is a Poissonian process with $p(t)=e^{-t/\tau_{jp}}$. For a
duration $2n_{p}/\Gamma_{m}$, the average number of switchings is
$\bar{n}_{jp}=2n_{p}/\Gamma_{m}\tau_{jp}$ with a deviation $\Delta n_{jp}=\sqrt{\bar{n}_{jp}}$.
The finite accuracy can be derived as \begin{equation}
\Delta V_{1}^{2}=\left|\frac{\partial\delta V}{\partial n_{jp}}\Delta n_{jp}\right|^{2}=\frac{4E_{z}^{2}+\Gamma_{m}^{2}}{8\Gamma_{m}T}\label{eq:dv1}\end{equation}
decreasing with the total measurement time $T$. Second, the variation
of the noise due to time evolution is $\Delta V_{2}^{2}=\sum_{n}\frac{\beta}{\omega_{n}}\langle\alpha_{n}^{2}\rangle\langle\sin^{2}(\omega_{n}T+\varphi_{n})\rangle\omega_{n}^{2}T^{2}$
for $B_{w}T\ll1$. With $\langle\alpha_{n}^{2}\rangle=1/3$ and $\langle\sin^{2}\varphi_{n}\rangle=1/2$,
we derive\[
\Delta V_{2}^{2}=\frac{\beta}{12\Delta\omega}B_{w}^{2}T^{2}\]
increasing quadratically with $T$. Hence the residue noise $\Delta V_{c}^{2}=\Delta V_{1}^{2}+\Delta V_{2}^{2}$
can be minimized by varying the measurement time. With our parameters,
the optimal time is $T\approx1400/\Gamma_{m}$.

\begin{figure}
\includegraphics[%
  bb=72bp 324bp 540bp 654bp,
  clip,
  width=7.5cm]{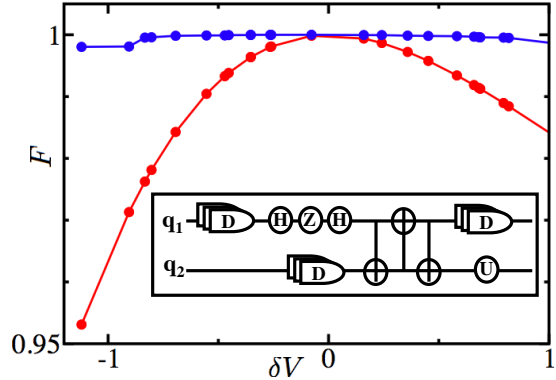}
\caption{\label{cap:fig3}Fidelity of the bit-flip gate. The solid circles
connected by the upper curve are the fidelity of the gate with the residue
noise; the solid circles connected by the lower curve are the fidelity
of the gate with the noise $\delta V(0)$. The box inside the plot:
the time sequence of the calibrations and the gates.}
\end{figure}

The calibration of the low-frequency noise can be exploited to improve
the fidelity of quantum-logic gates. Let a qubit be encoded with two
physical qubits, each of which is calibrated alternatively. The system
is programed so that at any moment one of the two physical qubits
has been calibrated and can be used for quantum memory or 
quantum-logic operations, while the other qubit is being calibrated. As an
example, we consider a bit-flip gate on a single qubit encoded with
physical qubits $q_{1}$ and $q_{2}$, as is shown in the box in Fig.\ref{cap:fig3}.
First, a continuous measurement is performed on qubit $q_{1}$ for
calibration. Afterward, the initial state is prepared in $q_{1}$
and the bit-flip gate is performed. At the same time, a continuous
measurement is performed on qubit $q_{2}$. After this calibration,
the state of $q_{1}$ is swapped to the calibrated $q_{2}$, which
is followed by other gate operations on $q_{2}$. Below we assume that
the off-diagonal coupling of our qubits can be tuned with large magnitude, while
the eigenenergies are nearly fixed  \cite{VionScience2002}.
In this case, a Hadamard gate $\hat{U}_{h}=e^{-i(\sigma_{z}+\sigma_{x})\frac{\pi}{2\sqrt{2}}}$
can be performed by applying an off-diagonal coupling $\hbar E_{z}$;
and a phase gate $\hat{U}_{ph}=e^{-i\sigma_{x}\frac{\pi}{2}}$ can
be performed by applying zero off-diagonal coupling. The bit-flip gate
is achieved with $\hat{U}_{flip}=\hat{U}_{h}\hat{U}_{ph}\hat{U}_{h}$.
The low-frequency noise adds a small off-diagonal term to the Hamiltonian
and affects the fidelity of the gates. The fidelity is defined as
$F=|\langle\psi_{t}|\hat{U}|\psi_{i}\rangle|^{2}$  \cite{NielsenBook}
where $|\psi_{i}\rangle$ is the initial state, $|\psi_{t}\rangle$
is the target state by ideal quantum gates, and $\hat{U}$ is the
gate operation with either the noise or the residue noise. In Fig.\ref{cap:fig3},
the fidelity of the bit-flip gate is plotted with the uncalibrated
noise and the residue noise respectively. Without calibration, the
fidelity decreases quadratically with the magnitude of the noise.
The calibration can significantly improve the gate performance to have
high fidelity in nearly all range of the noise. Note that this feature
is universal for other single-qubit gates and two qubit gates.

This approach of reducing decoherence due to the low-frequency noise
can be an important alternative to the dynamical control approach
and quantum error correction approach \cite{FacchiPRA2005}. In a dynamical
control approach, pulses much faster than $B_{w}^{-1}$ are applied
all the time during the gate operations and the quantum memory period
to cancel the effect of the noise \cite{ShiokawaPRA2004,GutmannPRA2005,FalciPRA2004,FaoroPRL2004}.
Meanwhile, it is necessary to carefully engineer the pulses to avoid affecting
the gate operations. In the quantum error correct approach,
many ancilla qubits are required to correct the quantum errors \cite{NielsenBook,Gottesman,KnillNature2005}.
While in our method, only a small number of physical qubits are needed
to encode one qubit. Between the calibrations, no extra pulse or measurement
is required. 

To conclude, we showed that the low-frequency noise can be calibrated
in the time-domain by a continuous measurement. Assisted
by the continuous measurement, the off-diagonal coupling of the low-frequency 
noise induces second-order relaxation between the qubit
states. We studied the stochastic evolution of the continuous measurement
by numerical simulation and used the switchings between the qubit
states to calibrate the noise. This approach can be a useful tool
for suppressing decoherence due to the low-frequency
noise in a solid-state qubit  in addition to existing approaches. The next steps are to study
the extension of this scheme to multiple qubits and to construct scalable
quantum computing protocols based on the calibration scheme.

\begin{acknowledgments}
We thank E. Knill and A. Shnirman for very helpful discussions and
the Qubit06 program of KITP for hospitality. 
\end{acknowledgments}

\end{document}